\documentclass{vldb}

\usepackage{cite}
\usepackage{booktabs} 
\usepackage{graphicx}
\usepackage{balance}
\usepackage{pifont} 
\usepackage{longtable}
\usepackage{enumitem} 
\usepackage[algoruled, linesnumbered, vlined]{algorithm2e}
\usepackage{algorithmic}
\usepackage{setspace}
\usepackage{multirow}
\usepackage{url}

\usepackage{sparklines}
\usepackage{xspace}
\usepackage{xcolor}
\usepackage{soul}
\usepackage{marginnote}

\newcommand{\ie}{i.e.,\ }
\newcommand{\eg}{e.g.,\ }

\newcommand{\tinyskip}{\vspace{3pt}}
\newcommand{\mypar}[1]{\tinyskip\noindent\textbf{#1.}\xspace}

\newcommand{\F}{\mbox{Fig.\hspace{0.25em}}}

\newenvironment{myitemize}{%
\begin{itemize}[leftmargin=1em, itemsep=.1em, parsep=.1em, topsep=.1em,
    partopsep=.1em]}
{\end{itemize}}

\newenvironment{myenumerate}{%
\begin{enumerate}[leftmargin=1em, itemsep=.1em, parsep=.1em, topsep=.1em,
    partopsep=.1em]}
{\end{enumerate}}

\newenvironment{structure*}{\color{blue}\begin{myenumerate}}{\end{myenumerate}}

\sethlcolor{yellow}

\vldbTitle{A Sample Proceedings of the VLDB Endowment Paper in LaTeX Format}
\vldbAuthors{Charles
Palmer, John Smith, Julius P.~Kumquat, and Ahmet Sacan}
\vldbDOI{https://doi.org/10.14778/xxxxxxx.xxxxxxx}
\vldbVolume{12}
\vldbNumber{xxx}
\vldbYear{2019}

\begin{document}

\title{Dataset-On-Demand: Automatic View Search \\ and Presentation for Data Discovery}

\numberofauthors{5}

\author{
\alignauthor 
Raul Castro Fernandez\\
\affaddr{The University of Chicago}\\
\email{raulcf@uchicago.edu}
\alignauthor
Nan Tang\\
\affaddr{QCRI}\\
\email{ntang@hbku.edu.qa}
\alignauthor
Mourad Ouzzani\\
\affaddr{QCRI}\\
\email{ouzzani@hbku.edu.qa}
\and
\alignauthor
Mike Stonebraker\\
\affaddr{MIT}\\
\email{stonebraker@csail.mit.edu}
\alignauthor
Sam Madden\\
\affaddr{MIT}\\
\email{madden@csail.mit.edu}
}

\maketitle

\begin{abstract}

Many data problems are solved when the right view of a combination of datasets
is identified. Finding such a view is challenging because of the many tables
spread across many databases, data lakes, and cloud storage in modern organizations.
Finding relevant tables, and identifying how to combine them is a difficult and
time-consuming process that hampers users' productivity.

In this paper, we describe Dataset-On-Demand (DoD), a system that lets users
specify the schema of the view they want, and have the system \emph{find} views
for them.  With many underlying data sources, the number of resulting views for
any given query is high, and the burden of choosing the right one is onerous to
users. DoD uses a new method, \emph{4C}, to reduce the size of the
view choice space for users. 4C classifies views into 4 classes:
\emph{compatible} views are exactly the same, \emph{contained} views present a
subsumption relationship, \emph{complementary} views are unionable, and
\emph{contradictory} views have incompatible values that indicate fundamental
differences between views. These 4 classes permit different presentation
strategies to reduce the total number of views a user must consider.

We evaluate DoD on two key metrics of interest: its ability to reduce the size
of the choice space, and the end to end performance. DoD finds all views within
minutes, and reduces the number of views presented to users by 2-10x.

\end{abstract}

\section{Introduction}
\label{sec:introduction}

Analysts of modern organizations need to answer questions that require combining
data from multiple different data sources, such as data lakes, databases and
even files in cloud storage. Example combinations of datasets, or \emph{views},
include training datasets necessary to build ML models, or the information
needed to answer a business question. Finding the desired view is hard for two
reasons. First, one must find relevant tables among many hundreds or thousands
of data sources in modern organizations.  Second, one must understand how to
combine the relevant tables to create the desired view. Most users do not know
how to find or join the view they need, often resorting to asking colleagues.
Even expert users find the process to be error-prone and time-consuming.

\begin{figure}[t]
  \centering
  \includegraphics[width=0.8\columnwidth]{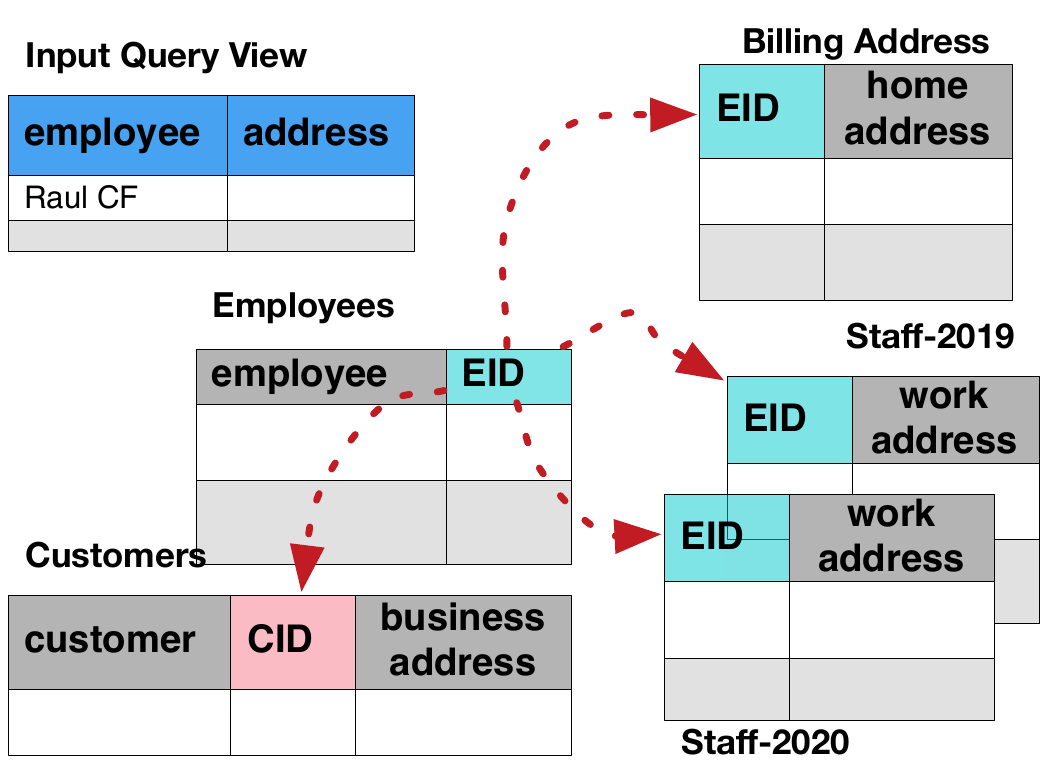}
\caption{Example of multiplicity of candidate views}
\label{fig:intro-example}
\end{figure}

Query-by-example~\cite{qbe,examplestutorial} based approaches such as
S4~\cite{sfour}, MWeaver~\cite{mweaver}, and Exemplar~\cite{exemplarqueries}
help non-expert users complete a partially fulfilled \emph{query view}.  The
query view may consist of attributes, tuple values, or a combination of both.
The systems identify a query that fulfills the query view.
However, they suffer two limitations.  First, they do not handle
semantic ambiguity. Consider the example of \F\ref{fig:intro-example}.  The
desired view is the top-left relation, which requests \textsf{employee} and
\textsf{address}.  It is possible to assemble 4 different views that fulfill
that specification by joining the \textsf{Employees} table with any of the
other tables. Joining with \textsf{Billing Address}, which contains \emph{home}
address, however, is semantically different than joining with
\textsf{Staff-2019} or \textsf{Staff-2020}, which contain \emph{work} address
and refer to two different years. In this case, the multiplicity of views stems
from semantic heterogeneity in the data, \ie multiple possible mappings, a
problem that has long been identified in the data integration community
\cite{schemamappingasquerydiscovery}. Second, these approaches assume knowledge
of the right join attributes between each pair of tables (they assume the
existence of a PKFK graph), so they always know the correct way of joining any
pair of tables. Assuming on the existence of a PFKF graph is a strong assumption we
do not make in this paper, because obtaining the right PKFK graph across many
tables stored in different systems and databases is a notoriously hard
problem~\cite{findpkfk1, findpkfk2}.  Instead of making such an assumption, we
concede that some join attributes will be wrong and deal with these cases. For
example, in \F\ref{fig:intro-example}, the \textsf{Employees} table may be
wrongly joined with \textsf{Customers} using \textsf{EID} and \textsf{CID} as
join attributes. Although there is an inclusion dependency between those
attributes (both are integers drawn from the same domain), it is nevertheless
not a valid join path, and will hence lead to a spurious view. We do not aim to
automatically detect these cases, but to identify them efficiently when
presenting views to users so they can select the correct one.

The crux of the problem is that the number of views that fulfill a given query
view is large due to semantic heterogeneity, erroneous join
attributes, and other anomalies. As a consequence, users are overwhelmed with
a large choice space, from where they have to select the view. 


In this paper we introduce dataset-on-demand (DoD), a \emph{view discovery
system} that helps users identify a combination of datasets (a view) from many
data sources. With DoD, users first declare the view they need by providing the
attributes they wish to see in the result as well as (possibly partial) tuples (see the
example query view in \F\ref{fig:intro-example}). Then, the system automatically
finds a collection of views that fulfill the specification. With DoD, finding
both the relevant tables and correct join attributes is treated as a single view
selection problem. DoD works by finding all possible views that fulfill the
specification, and then classifying those views into groups using a new method,
called \emph{4C}, which dramatically reduces the size of the view choice space.
As a result, users are only involved at the end of the process, and they choose
the right view from a smaller choice space instead of from all the resulting
views.

To reduce the size of the choice space, the 4C method first classifies all views
that fulfill a query view into 4 category groups, and then a \emph{presentation
strategy} uses those 4 groups to summarize, combine, and prune views.  The 4
category groups are: \emph{compatible}, which means that views are identical;
\emph{contained}, which means that the values of one view subsume others;
\emph{complementary}, which means that the set difference of tuples of views is
not empty; and \emph{contradictory}, which means that certain tuples are
incompatible across views. An example of contradictory views is shown in
\F\ref{fig:intro-example}: the view that selects \emph{home} address will
contradict the view that selects \emph{work} address, assuming the employee does
not live in the work place. Once the views are classified into the 4 groups, a
presentation strategy reduces the number of views users have to
inspect. There are different presentation strategies that cater to different
needs. For example, one presentation strategy may combine compatible views into
a single view, show only the highest cardinality view among contained views,
union complementary views, and ask users to pick the preferred view among
contradictory views. 

From a performance point of view, there are two operations that are expensive to
compute. First, classifying views into the 4 classes becomes expensive when
there are many input views. We propose a \textsf{4C-Chaining} algorithm that
minimizes the amount of computing resources needed by skipping computation when
possible.  Second, identifying all views involves searching for relevant tables
among many data sources, and obtaining all possible ways to join and materialize
them, which potentially involves executing a query across many data sources. We
build an ad-hoc processing engine to perform off-core joins when tables are part
of different data sources, \eg a database and a CSV file in cloud blob storage.
We build DoD on top of Aurum~\cite{aurum} to help with the search for relevant
tables and join paths.

To evaluate DoD we focus on two key metrics: reduction of the size of the view
choice space, and its end-to-end performance. We show that with 4C we
can reduce the size of the view choice space by 2-10X. We also demonstrate the
end to end performance of DoD, which can find and present views automatically
within minutes and in many cases within a few seconds. 


\section{Background}

In this section we present the problem dataset on demand solves
(Section~\ref{subsec:user_experience}) and how we evaluate it. We also describe
Aurum, on which we build DoD in Section~\ref{subsec:aurum}. The overall
architecture of DoD is shown in \F\ref{fig:arch}.

\mypar{Definitions} A query view, $qv$, consists of a set of attributes, $A$,
and, optionally, a set of tuples, $T$. A tuple is a set of values, $V$, drawn
from the attributes' domain. If $qv$ contains tuples, these can have values for
only a subset of $A$.  Each attribute and tuple value in a query view is called
an \emph{attribute} and \emph{value} constraint, respectively.

A view candidate, $c$, consists of a set of attributes $A' \in A$ and a set of
tuples $T'$ that contains the $T$ defined in $qv$.

\subsection{The Dataset-on-Demand Problem}
\label{subsec:user_experience}

Consider a user who wants to know what is the \textsf{address} of every
\textsf{employee} in a company. The user writes a query view with the relevant
attributes, as shown in the example of \F\ref{fig:intro-example}. We do not
assume the user knows the precise attribute names, but the DoD interface helps
with identifying them using an autocomplete feature. 

In addition to the schema, the user may add example tuples to the query view,
such as `Raul CF' in the example (see \F\ref{fig:intro-example}).  We assume
users may make mistakes when introducing example tuples, \eg typos. In addition,
data may be dirty and not match user's input. As a consequence, we do not expect
that values in the query view will appear exactly in the data.

\begin{figure}[t]
  \centering
  \includegraphics[width=0.8\columnwidth]{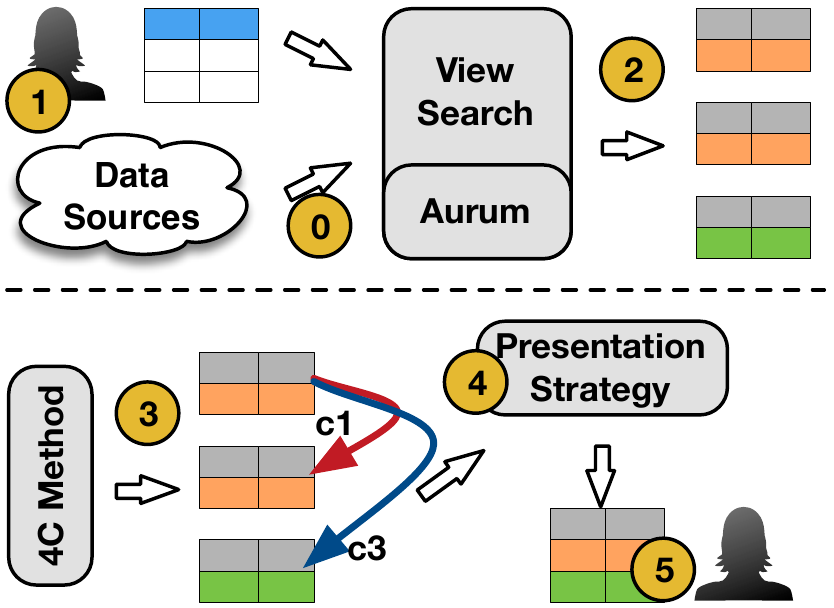}
\caption{Step 0: Aurum builds a discovery index from a collection of tables.
Step 1: A user creates and submits a query view. Step 2: DoD searches for views
based on the input query view. Step 3: The 4C method classifies all views into 4
groups (figure shows c1=compatible, and c3=complementary). Step 4: a
presentation strategy uses the 4 groups to reduce the number of views. Step 5:
the resulting views are presented to the user.}
\label{fig:arch}
\end{figure}

\mypar{View candidate search} Given $qv$, DoD finds a collection of
select-project-join queries that, when executed lead to a collection of
candidate views, $C$. In \F\ref{fig:arch}, step 1, the user submits a query view
to DoD, which produces (step 2) $C=3$ output views. In practice, the number of
views is often in the tens and more. The size of $C$ may be large for many
different reasons.  For example, because the attributes of the query view can
appear in multiple different tables, \eg `employee' may appear in many tables
that are irrelevant to this query. In addition, some of the inclusion
dependencies used to produce the views may be wrong, leading to spurious views
(such as joining with the table `Customers' in the example of
\F\ref{fig:intro-example}). Furthermore, some $c \in C$ may fulfill the query
view only partially, \eg lacking some attributes. These views have to be
considered as well, because if all views that fully fulfill the query view are
wrong, the users can consider these partially fulfilled views. We explain this
in Section~\ref{sec:engine}.

If DoD were to produce all $C$, the choice space users would need to investigate
would be very large: they would need to look at every view to choose the right
one, or settle with the first one they find that seems appropriate, at risk of
leaving better views unexplored. Although we can prioritize candidate views that
fulfill the largest number of constraints, there are still many to consider
manually, so we need a strategy to help users select the right candidate view. 

\mypar{View candidate selection} Given the candidate views, $C$, 4C classifies
them into 4 groups (step 3 of \F\ref{fig:arch}), and a \emph{presentation
strategy} is applied to reduce the total size of views that users must inspect.
This is shown in step 4 in \F\ref{fig:arch}.
Different presentation strategies cater to different needs. For example, with
the \textsf{4c-summary} strategy (explained later) users may interact with the
system indicating `good' views and `bad' views, and this information can be used
to further narrow down the search. Alternatively, DoD can avoid any user
intervention with a \textsf{multi-row relation} presentation strategy we present
later. We explain this in Section~\ref{sec:4c}.

\underline{\emph{4C Example:}} Suppose the \emph{`employee-address'} query view
above produces 16 views.  Out of those, 14 fully fulfill the query view, and 2
others are partial, \ie lack some attribute defined in the query view. Because
we prefer views that fully fulfill the query view, we apply the 4C method with
the \textsf{4C-summary} (presented more in detail later) presentation strategy
to the group with 14 views first.  The 4C method finds that 8 of the 14 views
are \emph{compatible}, \ie they are exactly the same. This can happen, for
example, because equivalent join attributes were used to assemble the views, or
because there are copies of the underlying tables in different databases. In
this case, \textsf{4C-summary} summarizes those 8 views into 1. Out of the
remaining 6 views, 3 of them are \emph{contained} in another one. This can
happen, for example, because a view joined with tables that contained employees
of \emph{engineering}, and another view joined with tables that contained both
\emph{engineering} and \emph{sales}. Instead of showing the 4 views,
\textsf{4c-summary} shows only the one that contains the other 3. 

Suppose the previous two summarized views are \emph{complementary}: each view
has values that do not exist in the other one. This can happen due to the
existence of different versions of the same table. \textsf{4c-summary} unions
these two views into 1.  Finally, out of the remaining views, 4C finds a pair
that is \emph{contradictory}, that is, when looking at a particular key in one
view, the row values are different than when looking at that same key in the
other view. For example, an employee `Raul' has a `Pie street' address in one
view, but a `Flea Av' in another one. This can happen because the address
attribute in one view referred to `work address', while another one to `home
address'.  It is possible to ask users to resolve contradictions, and use their
actions to further reduce the size of the choice space.  Alternatively, another
presentation strategy can assemble a single view with multiple values for
employee `Raul', so this contradiction can be dealt with downstream. At the end,
instead of looking at 14 different views, 4C with \textsf{4c-summary} reduces
the size to a few choices, so it becomes much easier to select the right view
(step 5 in \F\ref{fig:arch}).

Classifying all the views in $C$ into the 4 classes is an expensive
process because it is necessary to compare all cells of each view to all other
views. We show later the algorithms we introduce to make the process practical.


\subsection{Data Discovery in Databases, Data Lakes, and Clouds with Aurum}
\label{subsec:aurum}

In this section, we describe Aurum~\cite{aurum}, an open source data discovery
platform on top of which we build DoD.  


\mypar{Aurum overview} Aurum reads relations from data sources such as
databases, lakes, and files in filesystems, and produces a \emph{discovery
index} (corresponds to step 0 in \F\ref{fig:arch}). The discovery index is
logically a hypergraph where the nodes correspond to columns and edges represent
relationships between columns, \eg there is an inclusion dependency
relationship. Hyperedges are used to indicate columns that are part of some
hierarchy, \eg they appear in the same relation.  This is useful to identify
tables from relevant columns. In addition to these edges, the discovery index
also includes a full text search index of textual columns, as well as their
attribute names.

Aurum provides a discovery API that uses the discovery index to answer complex
discovery queries. For example, a discovery program can ask for nodes in the
graph (columns) that contain a value, and then it can ask whether the tables
that contain those columns (using hyperedges) can be joined together. We detail
the relevant discovery API calls when describing DoD's engine in the next
section.

\mypar{Why approximate inclusion dependencies suffice} State of the art
approaches assume that a PKFK graph that indicates how to
join every pair of tables is available \cite{sfour, mweaver, exemplarqueries}.
This is a strong assumption we do not make. Obtaining such a graph across
multiple databases is extremely hard, and even if it existed, it would not be
enough to solve the problem of semantically different join paths (see the
example in Section~\ref{sec:introduction}).

DoD instead relies on inclusion dependencies, which Aurum finds automatically,
and uses these as a proxy to understand which tables join with each other on
which attributes. When using inclusion dependencies, we accept that some will not
be correct attributes on which to join tables, and will therefore lead to wrong
views. Nevertheless, this approach works because spurious views are classified
and contrasted with other (correct) views as part of the 4C method, so users can
choose the correct one.

Further, instead of using exact inclusion dependencies, DoD uses
\emph{approximate} inclusion dependencies. An approximate inclusion dependency
between two columns exists when values of one column are approximately included
in the other, and at least one of the two columns is almost unique. These
relaxations help with two practical problems. First, it is sometimes useful to
use approximate keys to combine datasets that would not be able to be combined
otherwise, \eg joining two tables on an attribute `Office Phone' that is not a
real key. This is true when tables come from different sources and nobody
assigned a PKFK at design time. Second, because data is dirty, if we only
considered exact keys, we'd ignore 
opportunities to combine datasets.

\subsection{Performance Metrics}

There are two key metrics to evaluate DoD: the ability of 4C to reduce the total
number of views,
and the end-to-end runtime. To measure the first, we
apply the presentation strategy and measure how many views remain at the end. 

Second, if finding and classifying all views takes hours, then the benefits of
DoD in comparison to manually inspecting all views would be unclear, so it is
important for the end to end runtime to be not more than a few minutes.

\section{The 4C Method}
\label{sec:4c}

We assume we have found all candidate views, $C$, (step 2 in \F\ref{fig:arch}),
and present the 4C method. 4C classifies views into four classes
that we use to reduce the size of $C$. We start formally defining the 4
categories in Section~\ref{subsec:4c}. Then we describe the algorithms
(Section~\ref{subsec:algo}) and conclude the section describing different
presentation strategies (Section~\ref{subsec:presentation}).

\subsection{Compatible, Contained, Complementary, Contradictory Views}
\label{subsec:4c}

A candidate view, $v_i$ contains a set of rows, $R_i$, that can be retrieved with the
function $f$, $f(v_i) = R_i$. We can refer to a row, $r_i \in R_i$, with its key value,
$k_i$, which we can use to obtain the row; $k(R_i, k_i) = r_i$.

\mypar{Compatible views} Two candidate views, $v_1$ and $v_2$, are compatible if
they have the same cardinality $|R_1| = |R_2|$, and their set of tuples is
exactly the same: $|(R_1 \setminus R_2)| = 0$ and $|(R_2 \setminus R_1)| = 0$,
where the symbol $\setminus$ indicates the set difference operation. 

\mypar{Contained views} A view, $v_1$, contains another view, $v_2$, when $|(R_1
\setminus R_2)| > 0$ and $|(R_2 \setminus R_1)| = 0$, that is, when all rows of
$v_2$ are contained in $v_1$. 

\mypar{Complementary views} Two views, $v_1$ and $v_2$ are complementary when
$|(R_1 \setminus R_2)| > 0$ and $|(R_2 \setminus R_1)| > 0$, that is, each view
has rows not contained in the other view.

\mypar{Contradictory views} A view, $v_1$ contradicts another view, $v_2$, if a
key value, $k_i$, that exists in both views yields distinct rows, $k(R_1, k_i)
\ne k(R_2, k_i)$. Two rows are distinct when any value of any of their
attributes is different.

\smallskip

Although the description above refers to two views, the 4C method classifies
multiple views into each of the above classes. For example, many views can be
compatible among themselves. One view may contain many other views. One single
view can be complementary with respect to many others, and contradict many
others as well.  Similarly, a view can contain another view, be complementary to
another one, compatible with another group and be in contradiction with other
views on different rows, all at the same time. Next, we present the algorithm we
use to classify the candidate views into the 4 above groups.

\subsection{The 4C-Chasing Algorithm}
\label{subsec:algo}

Obvious methods to obtain the 4C classification of views perform poorly. For
example, an algorithm could compare each cell of each view to all other views,
but this would become too slow even when the number of views is low. One good
baseline method hashes rows of views and use the set of hashes to
quickly tell apart compatible and contained groups from complementary and
contradictory. Unfortunately, this improved baseline still needs to use the
per-cell comparison to distinguish between complementary and contradictory
views, so its performance remains low.

We introduce the \textsf{4C-Chasing} algorithm to speed up the classification.
This algorithm is orders of magnitude faster than the baseline, as we show in
the evaluation section. 

\subsubsection{Core 4C-Chasing}

The main body of the algorithm is shown in Algorithm~\ref{alg:main}. 
We know each $c \in C$ has a subset of the
attributes defined in the query view. When the candidate view contains all
attributes from the query view, we say their schema fully fulfills the
query view. We prefer these views because they are closer to the definition given by
the user. However, we must also consider candidate views with a schema that
partially fulfills the query view (\ie they contain only a subset of the
attributes in the query view), in case all fully fulfilled views may be wrong.

Because 4C works on views with the same schema (attributes in this case), the
first step is to separate the candidate views into groups according to their
schema (line~\ref{alg:4c:classifyschema}).  Then, the algorithm runs the 4C
classification for each group (see line~\ref{alg:4c:foreachschema:start}),
leading to the classification of views into the four groups: compatible
(\textsf{C1}), contained (\textsf{C2}), complementary (\textsf{C3}), and
contradictory, \textsf{C4}.


\begin{algorithm}[ht]
  \small
  \begin{spacing}{0.8}
      \caption{Main 4C Algorithm}
  \label{alg:main4c}
  \SetKwInOut{Input}{input}%
  \SetKwInOut{Output}{output}%
  \Input{\ $V$, collection of candidate views
    } 
  \Output{\ $G$, views classified per class and schema type
    }

  \BlankLine
  $S$ $\leftarrow$ classify\_per\_schema($V$)\label{alg:4c:classifyschema}\;
  \For{$s \in S$}{\label{alg:4c:foreachschema:start}
    $V$ $\leftarrow$ find\_candidate\_keys($V$)\;\label{alg:4c:metadata}
    $C1$ $\leftarrow$ identify\_c1($V$)\;\label{alg:4c:compatible}
    $SC1$ $\leftarrow$ select\_representative\_c1($C1$)\;\label{alg:4c:selection}
    $(C2, C34)$ $\leftarrow$ identify\_c2\_and\_candidate\_c3c4($SC1$)\;\label{alg:4c:c2}
    $(C3, C4)$ $\leftarrow$ identify\_c3\_and\_c4($C34$)\;\label{alg:4c:c34}
    $G[s]$ $\leftarrow$ $(C1, C2, C3, C4)$
  }
  \BlankLine
  \Return $G$\;
\end{spacing}
\label{alg:main}
\end{algorithm}


\begin{algorithm}[ht]
  \small
  \begin{spacing}{0.8}
      \caption{identify\_c1}
  \label{alg:identifyc1}
  \SetKwInOut{Input}{input}%
  \SetKwInOut{Output}{output}%
  \Input{\ $V$, collection of candidate views
    } 
  \Output{\ $C1$, collection of lists, each list contains $v \in V$ that are
compatible
    }
  
  \BlankLine
  $g$ $\leftarrow$ map()\;
  \For{$v_i \in V$}{\label{alg:c1:foreachview}
    $hv_i$ $\leftarrow$ hash($v_i$)\;\label{alg:c1:hashview}
    add($g$, $hv_i$, $v_i$)\;\label{alg:c1:groupby}
  }
  $C1$ $\leftarrow$ $g.values()$\;
  \Return $C1$\;

\end{spacing}
\label{alg:c1}
\end{algorithm}


\begin{algorithm}[ht]
  \small
  \begin{spacing}{0.8}
      \caption{identify\_c2\_and\_candidate\_c3c4}
  \label{alg:identifyc2}
  \SetKwInOut{Input}{input}%
  \SetKwInOut{Output}{output}%
  \Input{\ $SC1$, collection of candidate views
    } 
  \Output{\ $C2$, list of tuples $(v_C, [v_c])$; $v_C$ contains all $[v_c]$\\
          \ $C34$, list of tuples $(v_i, v_j, I_i, I_j)$; $I_{i,j}$ are
the sets of row indices in which $v_i$ and $v_j$ disagree 
    }
  
  \BlankLine
  $(gc2, gc34)$ $\leftarrow$ (map(), map())\; 
  \For{$(v_i, v_j) \in pairs(SC1)$}{\label{alg:c2:foreachview}
    $(h\_list_i, h\_list_j)$ $\leftarrow$ (hash\_rows($v_i$), hash\_rows($v_j$))\;
    $(h\_set_i, h\_set_j)$ $\leftarrow$ (set($h\_list_i$), set($h\_list_j$))\;
    \If{len($h\_set_i$) $\ge$ len($h\_set_j$)}{\label{alg:c2:startc2}
      \If{$(h\_set_j \setminus h\_set_i) = 0 $}{
        add($gc2$, $v_i$, $v_j$)\;\label{alg:c2:endc2}
      }
    }
    \Else{\label{alg:c2:startc34}
      $s_{ij}$ $\leftarrow$ $h\_set_i \setminus h\_set_j$\;
      $s_{ji}$ $\leftarrow$ $h\_set_j \setminus h\_set_i$\;
      \If{$s_{ij} \ge 0 \wedge s_{ji} \ge 0$}{
        $I_i$ $\leftarrow$ indices($h\_list_i$, $s_{ij}$)\;
        $I_j$ $\leftarrow$ indices($h\_list_j$, $s_{ji}$)\;
        add($gc34$, $v_i$, $v_j$, $I_i$, $I_j$)\;\label{alg:c2:endc34}
      }
    }
  }
  $(C2, C34)$ $\leftarrow$ (unpack($gc2$), unpack($gc34$))\;
  \Return $(C2, C34)$\;

\end{spacing}
\label{alg:c2}
\end{algorithm}


\begin{algorithm}[ht]
  \small
  \begin{spacing}{0.8}
      \caption{identify\_c3\_and\_c4}
  \label{alg:identifyc2}
  \SetKwInOut{Input}{input}%
  \SetKwInOut{Output}{output}%
  \Input{\ $C34$, list of tuples $(v_i, v_j, I_i, I_j)$; $I_{i,j}$ are
the sets of row indices in which $v_i$ and $v_j$ disagree 
    } 
  \Output{\ $C3$, list of tuples $(v_i, v_j, I_i, I_j)$; $I_{i,j}$ are
the sets of row indices that are existing in $i$ (or $j$) and not the other view\\
          \ $C4$, list of tuples $(v_i, v_j, k, K_i)$; $K_i$ are the
sets of key values (attribute $k$) in which $v_i$ presents a contradiction with
respect to $v_j$ 
    }
  
  \BlankLine
  $G$ $\leftarrow$ build\_graph($C34$)\;\label{alg:c34:buildgraph}
  seen $\leftarrow$ set()\;
  $(C3, C4)$ $\leftarrow$ (set(), set())\;
  \While{$C34 \neq \{\}$}{
    $c34$ $\leftarrow$ $C34$.pop()\;\label{alg:c34:pop}
    \If{$c34$ in seen}{
      continue\;
    }
    $vsel_i$ $\leftarrow$ select($c34.v_i$, $c34.I_i$)\;\label{alg:c34:sel1}
    $vsel_j$ $\leftarrow$ select($c34.v_j$, $c34.I_j$)\;\label{alg:c34:sel2}
    $k$ $\leftarrow$ get\_most\_likely\_key($c34$)\;\label{alg:c34:findkey}
    $KV_i$ $\leftarrow$ project($vsel_i$, $k$)\;\label{alg:c34:proj1}
    $KV_j$ $\leftarrow$ project($vsel_j$, $k$)\;\label{alg:c34:proj2}
    $KV_{ix}$ $\leftarrow$ $KV_i \cap KV_j$\;\label{alg:c34:contkeys}
    $kv$ $\leftarrow$ $KV_{ix}$.pop()\;\label{alg:c34:digstart}
    $k_c, V_c$ $\leftarrow$ find\_contrad($vsel_i$, $vsel_j$, $k$, $kv$)\;
    mark\_graph\_node($G$, $c34$, $(k_c, V_c)$)\;
    add($C3$, $c34$)\;
    add(seen, $c34$)\;\label{alg:c34:digend}
    $CK$ $\leftarrow$ $(KV_i \cup KV_j) \setminus KV_{ix}$\;\label{alg:c34:compkeys}
    add($C4$, $c34$)\;
    add(seen, $c34$)\;
    \While{contains\_marked\_nodes($G$)}{\label{alg:c34:chasestart}
      $(c34, k_c, V_c)$ $\leftarrow$ obtain\_marked\_node($G$)\;
      \For{neighbor $\in$ $c34.v_i$}{
        $k_c, V_c$ $\leftarrow$ find\_contrad($c34.v_i$, neighbor, $k$, $kv$)\;
        \If{$(k_c, V_c)$}{
          mark\_graph\_node($G$, $c34$, $(k_c, V_c)$)\;
          add($C3$, $c34$)\;
          add(seen, $c34$)\;\label{alg:c34:chaseend}
        } 
      }
    }
  }
  \Return $(C3, C4)$\;

\end{spacing}
\label{alg:c34}
\end{algorithm}

The algorithm first identifies quickly compatible and contained groups of views.
Then, it identifies complementary and contradictory views among the remaining
ones.

\mypar{Prepare views} In the main body of the loop, the algorithm first attaches
metadata to each view, which consists of a value per attribute that indicates
how likely the attribute is to be a key (line~\ref{alg:4c:metadata}). This is
computed as the ratio between the total number of distinct values in the
attribute column and the total number of values in it. This metadata
will be used later in the \textsf{identify\_c3\_and\_c4()} function. 

\mypar{Identify compatible groups} The next step is to identify groups of views
that are compatible (c1) with each other (line~\ref{alg:4c:compatible}), which
corresponds to Algorithm~\ref{alg:c1}. The main idea is to
hash each view (see line~\ref{alg:c1:hashview}), and insert
the view on a map that is keyed on the hash value. A view hash is the sum of its
rows hashes, and a row hash is the sum of its cells hashes.  With this hashing
method, compatible views are guaranteed to have the same hash value, so this is
an efficient way of identifying groups of compatible views fast. 

Because all compatible views are identical, we select one only to continue the
classification (line~\ref{alg:4c:selection}).
The selected views are then given to a function in charge of identifying contained
groups (c2), as well as groups that may be either complementary or contradictory
(c34), see line~\ref{alg:4c:c2}. 

\mypar{Identify contained groups} The function to finding contained views is
shown in Algorithm~\ref{alg:c2}. For each pair of views (loop in
line~\ref{alg:c2:foreachview}), the algorithm obtains the list and the set of
hashes of each view respectively. The list will be used to identify indexes of
rows (because it maintains the order), while the sets are used to quickly tell
whether the views are contained or may be complementary.
Lines~\ref{alg:c2:startc2} to \ref{alg:c2:endc2} checks whether the hashes of
one view are completely contained on the first, in which case the view is
contained in the first and this is recorded in the variable $gc2$. Note that
because the loop enumerates pairs of views in both directions, it is not
necessary to check for containment in both directions here. 

If the views are not contained, the algorithm tries to decide if they are
candidates for the complementary and contradictory group. This decision 
is done in lines~\ref{alg:c2:startc34} and \ref{alg:c2:endc34}. The
algorithm computes the set difference of rows in both directions and checks
whether it is non-empty in both cases. When it is, this indicates that each view
has hashes not contained in the other view. This can be caused for two different
reasons: either one view has rows that are non existent in the other (a case of
complementarity), or the values of certain rows differ in some values, which
would also lead to different hashes and indicate a contradiction. The algorithm
returns the identified groups of contained views as well as those that may be
complementary or contradictory.

\subsubsection{Distinguising Complementary and Contradictory Views}

The last part of the algorithm is to identify which pairs of tables in $C34$ are
complementary and which are contradictory.  This corresponds to
line~\ref{alg:4c:c34} in the main body of algorithm~\ref{alg:main}. The body of
this function is shown in more detail in Algorithm~\ref{alg:c34}. 

Checking whether views are complementary or contradictory is expensive. This is
because it's not possible to recover the view row values from the hash. Hence,
it requires checking each row cell individually for all rows that are part of
the set difference, and comparing it with all other views.

The insight that the 4C-chasing algorithm exploits is that if there exists a
particular contradiction in one value between 2 views, it is likely that at
least one of the views contradicts other views on the same value as well---for
example, all views that were assembled using a particular join attribute between
two tables will share the same contradictions. Therefore, instead of finding
contradictions between each pair of views, we can use the contradictions we find
to quickly test whether the same contradiction exists between other pairs too,
\ie we can chase the contradiction across the views in the group.

\mypar{Separate complementary from contradictory} The algorithm first represents
the pairs of views in an undirected graph (the chasing graph), see
line~\ref{alg:c34:buildgraph}. It then obtains a pair from the input pairs
(line~\ref{alg:c34:pop}) and checks whether the different values (hashes)
identified between this pair are due to complementarity or contradictions. For
that, it first selects the relevant rows from each view
(lines~\ref{alg:c34:sel1} and \ref{alg:c34:sel2}), and then it obtains the
values for the key of each of those views. The key values are obtained by first
identifying the most likely attribute key, $k$, of those views
(line~\ref{alg:c34:findkey}), and then projecting that attribute on the
selection (lines~\ref{alg:c34:proj1} and \ref{alg:c34:proj2}). This produces
$KV_i$ and $KV_j$, which are the key values that correspond to the distinct
hashes of each view. At this point, the intersection of these two sets,
$KV_{ix}$, corresponds to the contradictory examples, while the rest of values
correspond to the complementary ones (line~\ref{alg:c34:compkeys}).  When the
keys of two views are contradictory, the views will be classified as
complementary because the set of hashes will be different, and the key
intersection, $KV_{ix}$, is empty. 

Within the set of contradictory keys, it is still necessary to identify which
cell value or values caused the contradiction. We want the specific cell values
so we can test these with other views, and so we can include this fine-grained
information with the metadata produced at the output. The contradictory cells
are obtained in lines~\ref{alg:c34:digstart} to \ref{alg:c34:digend}. Here we
call \textsf{mark\_graph\_node()} on the graph as contradictory, and attach the
key attribute, $k_c$, (note that $k_c$ does not need to be the same as $k$) as
well as the key values that produced a contradiction, $V_c$.

\mypar{Chasing graph} The last part of the algorithm uses the nodes of the graph
which has been previously marked (due to a contradiction), and the already found
contradiction to identify what other connected views contradict the node as well
(lines~\ref{alg:c34:chasestart} to \ref{alg:c34:chaseend}).  Without
\textsf{4C-chasing}, testing a pair involves performing $r$ operations in the
worst case, with $r$ being the number of rows in the larger cardinality
relation. This cost must then be paid for all-pairs of contradicting views,
$|c34|^2$. In contrast, with the new algorithm, the cost of comparing each pair
is only $1$ for cases where the contradictions are shared. The benefits of the
algorithm are then directly related to how many contradictions are shared
between views. Since contradictions are often shared across many views 
this algorithm is efficient.




Once finished, all $C$ are classified into 4 classes, which along with the query
used to assemble the view, and the join graphs used, is given as input to a
\emph{presentation strategy}.
%
Specialized users such as data
stewards may want a fine-grain understanding of the SPJ used to assemble a
specific view, or the indices of rows in which two views disagree, or
about the attribute and value that correspond to the contradiction between a
pair of views. They can use the information collected so far directly.

\subsection{Presentation Strategies}
\label{subsec:presentation}

The goal of a presentation strategy is to use the 4 classes into which the
candidate views are classified in order to reduce the size of the view
choice space. We explain next two such strategies:

\mypar{4c-summary} This strategy reduces the view choice space by taking
automatic actions for compatible, contained, and complementary views, and then
asking a user to choose an option among contradictory views. This strategy was
illustrated in the example in Section~\ref{subsec:user_experience}. It
summarizes compatible groups with a representative view, and contained groups with the
highest cardinality view. When views are complementary, the union of the views
is shown. Finally, when views are contradictory, the system shows to users the
view that contradicts the largest number of views. If users prefer one view 
the information is used to prune other views, otherwise, the next contradiction
is shown. This strategy makes two key assumptions. First, that when views are
contained, the highest cardinality one is preferred. Second, that when views are
complementary, the union is preferred. These actions do not guarantee that the
view the user wants is chosen, \eg a user may prefer the view for a particular
year (that may be contained in another view). However, users can always inspect
the operations performed by the presentation strategy and recover the right
view.

\mypar{Multi-row views} This strategy does not require user intervention.  A
view is shown exactly only once for each schema group. When a view is both in a
compatible, contained, and complementary group, it is unioned with its
complementary views and shown once. When the view has contradictory views, the
system performs a special union, where those rows that contradict each other
become \emph{multi-rows}. A multi-row is a tuple that for the same key has more
than one value. This is not a valid relation, but it can be useful for users to
have all the context to make their decision, and it can be useful to feed to
certain downstream tasks---for example one that uses multi-rows to apply an
entity resolution \cite{entityresolution} algorithm.

\smallskip

Different presentation strategies cater different needs and present different
tradeoffs, and more presentation strategies than the presented above are
possible. While \textsf{4c-summary} relies on minimal user intervention to deal
with contradictory views, the \textsf{multi-row} strategy further automates the
process. In all cases, all metadata collected is shown along with the view so
users can build trust on the results.

\section{DoD Engine}
\label{sec:engine}

In this section we describe how DoD finds $C$, which corresponds to step 2 in
\F\ref{fig:arch}. 
We start by describing the search process in Section
\ref{subsec:assembly-process}, as well as some details on the ad-hoc processing
engine we use to join relations across sources
(Section~\ref{subsubsec:ismaterializable}).  

\subsection{Finding Candidate Views}
\label{subsec:assembly-process}

The search process finds $C$ from an input query view
following the next stages:

\subsubsection{Identify Candidate Tables}

During the first step, DoD identifies the tables
\emph{relevant} to the input query view. A table is relevant to the query
view if it fulfills an attribute constraint, a value constraint, or
any number of the previous. A table fulfills an attribute constraint when it
contains one attribute that appears in the query view. A table fulfills a
value constraint when the table contains both an attribute and value
constraint that were defined in the same column of the query view. Note that a
table that only fulfills a value constraint, but not the corresponding attribute
constraint, is not considered relevant: this is because values alone can appear
in many different columns of the underlying datasets. 

To identify underlying tables that fulfill an attribute constraint, DoD uses
Aurum's data discovery API. Given an attribute name, Aurum returns all columns
(and the corresponding tables) that have an attribute name that matches (string
equality) the input attribute name.  Because users make use of Aurum's
autocomplete feature to assemble the attributes in the query view in the first
place (see Section~\ref{subsec:user_experience}, exact string matching is sufficient to
identify the tables.

To identify underlying tables that fulfill a value constraint, DoD
uses Aurum again. It first finds the set of tables that fulfill the attribute
constraint, as above. The results are the attribute constraint group. Then, DoD
uses Aurum to search for the value constraint. Aurum returns columns that
contain the value constraint using a full-text search index.  This is the
value constraint group. Finally, DoD takes the set intersection of both
groups: the result is the set of tables that fulfill the value
constraints.

The above procedure is performed for every attribute and value
constraint in the query view. The output of this process is a list of
\emph{relevant} tables along with the query view constraints they satisfy. We
call the union of relevant tables the \emph{candidate tables}.

\subsubsection{Find Candidate Groups}

In this step, DoD wants to identify a group of tables that, if joined, would
satisfy the query view definition. The ideal group would fulfill all the
constraints in the query view, and would contain as few tables as possible---so
fewer joins would be necessary. Finding this group is akin to solving a set
cover problem over the candidate tables. This, however, is not sufficient,
because it may be impossible to join the tables in that group.

Instead, DoD finds \emph{all} subsets of tables from the candidate tables set
that, together, fulfill as many constrains from the input query view as
possible. Each subset of candidate tables is called a \emph{candidate group}.
Each candidate group, if materialized, would lead to a view that fulfills or
partially fulfills the query view.

DoD performs the search of candidate groups using a procedure that caters to two
preferences. First, groups that fulfill more constraints are preferred, as those
are closer to the query view. Second, among candidate groups that fulfill the
same number of constraints, smaller groups are preferred, because those involve
fewer joins. These criteria is implemented in a greedy search process that is
shown in detail in Algorithm~\ref{alg:candidatetables} and works as follows:

\begin{algorithm}[ht]
  \small
  \begin{spacing}{0.8}
      \caption{Find candidate groups}
  \label{alg:main4c}
  \SetKwInOut{Input}{input}%
  \SetKwInOut{Output}{output}%
  \Input{\ $T$, collection of relevant tables and the constraints they fulfill
    } 
  \Output{\ $G$, list of candidate groups
    }

  \BlankLine
  $G$ $\leftarrow$ []\;
  c\_group $\leftarrow$ []\;
  c\_group\_constraints $\leftarrow$ []\;
  stables $\leftarrow$ sort\_tables\_number\_constraints($T$)\;\label{alg:ct:sor}
  \For{ref$ \in $stables}{\label{alg:ct:s}
    c\_group.add(ref)\;
    ref\_contraints $\leftarrow$ get\_constraints(ref)\;
    c\_group\_constraints $\leftarrow$ ref\_constraints\;\label{alg:ct:e}
    \If{fulfill\_query\_view(c\_group\_constraints)}{\label{alg:ct:check1}      
      $G$.add(c\_group)\;
      continue\;
    }    
    \For{t $\in$ tables\_after\_ref(ref)}{\label{alg:ct:so}
      constraints $\leftarrow$ get\_constraints(t)\;
      new\_constraints $\leftarrow$ merge(ref\_constraints, constraints)\;
      \If{$|$new\_constraints$| \ge 0$}{\label{alg:ct:check}
        c\_group.add(t)\;
        c\_group\_constraints.add(constraints)\;
        \If{fulfill\_query\_view(constraints)}{\label{alg:ct:check2}
          c\_group.add(t)\;
          $G$.add(c\_group)\;
        }
      }
    }
  }
  \BlankLine
  \Return $G$\;\label{alg:ct:done}
\end{spacing}
\label{alg:candidatetables}
\end{algorithm}

DoD sorts the candidate tables based on the number of constraints they fulfill
(line~\ref{alg:ct:sor}), from higher to lower. The search process takes the first
table of such list, called reference (\textsf{ref} in the algorithm) and adds it
to a candidate group (lines~\ref{alg:ct:s}-\ref{alg:ct:e}). Then, it iterates
over the tables in sort order (line~\ref{alg:ct:so}) checking, at each step,
whether the table fulfills constraints not covered by the reference table
(line~\ref{alg:ct:check}). When it does, the table is added to the candidate
group. At this point, if all constraints are fulfilled by the candidate group,
the group is stored and the process selects a new reference table. If not, then
the search continues. Note that in some cases, a single table may fulfill all
constraints (that's why it's necessary to check for query view fulfillment every
time the candidate group is updated, such as in lines~\ref{alg:ct:check1} and
\ref{alg:ct:check2}. This happens when the input query view corresponds to an existing
table.

The output of this step is a collection of candidate groups
(line~\ref{alg:ct:done}), each with the set of constrains it fulfills. 

\subsubsection{Select Joinable Groups}
\label{subsubsec:joinable}

The goal of this step is to select among the candidate groups, those that are
\emph{joinable groups}. A joinable group is a candidate group in which all the
tables contained can be combined into one single table through the relational
join operation. There may be multiple ways of joining the tables of a joinable
group. This can happen if there exists more than one inclusion dependency
between a pair of tables, or there are different intermediate tables that can be
used to join two tables. Each strategy to join such tables is called a
\emph{join graph}. A joinable group, then, is a candidate group for which there
is at least one existing join graph. 

For each candidate group, DoD must find all join graphs that permit combining
all tables. It is necessary to find all join graphs because some may be using a
wrong inclusion dependency (leading to a wrong view) or one that is correct, but
incompatible with the user's intent---when there are different semantic views
that fulfill the query view.

To find all the join graphs, DoD first finds all the join paths between every
pair of tables in the group. Then, it combines join paths together to form join
graphs that permit joining all tables in the group. DoD discards duplicate join
graphs---those that join the same tables on the same attributes---as well as
join graphs that do not join every table in the group.

In order to identify a join path between a pair of tables, DoD uses Aurum's
discovery API. Aurum returns all join paths between two tables, given a maximum
number of hops, by querying its discovery index. Once all join graphs are
identified, these are sorted based on the number of joins it is necessary to
perform in each case. The output of the process is then the joinable groups,
each with the fulfilled query view constraints.

\mypar{Note on value constraints} A joinable group indicates that when the
tables are joined together, the materialization will contain some of the
attribute constraints defined in the query view. It cannot say, however, whether
the materialization will contain a value constraint. This is because Aurum can
identify whether there are inclusion dependencies between two tables, but it
cannot answer whether the tables, when joined, would contain a particular value.
Identifying what joinable groups will contain the desired specification at the
output is the goal of the next step.

\subsection{Select Materializable Groups}
\label{subsubsec:ismaterializable}

A materializable group is a joinable group that, when materialized, contains all
the constraints of the joinable group, including the value constraints. Because
Aurum cannot help with identifying materializable groups, DoD checks each
joinable group individually, \ie it performs the join and checks whether the
value constraint appears at the output.



In this section, we discuss alternative ways of achieving that, and explain the
alternative we chose. We then dedicate the rest of the section to specific
optimizations we had to implement in DoD to increase its efficiency.

The need for querying across data sources calls for a federated query
engine~\cite{haasfederation, bigdawg}. These engines receive queries expressed
in some general language, \eg SQL, and compile them into specific languages of
the data sources. They have a couple of drawbacks. First, such engines require
to configure a \emph{connector} for each data source and code the logic to
transform from the general query language to the specific data source one.
Second, they need to interact with all the data sources in the enterprise,
making these systems hard to deploy in practice.

Another alternative would be to assume all data is in a data lake with querying
capacity, and run all queries there directly. However, in practice we have
found that many times data must remain in certain data sources (\eg for security
or governance issues), or has not arrived to the lake yet. We only assume read
access.

We implemented an ad-hoc query engine that does not suffer the deployment
problems of federated systems and does not assume all data is in a central
repository, or data lake. The engine is implemented using Python
Pandas~\cite{pandas}. Its main advantage is that instead of relying on external
querying capacity, it owns query processing. Despite its slower performance with
respect to query engines of databases, the benefits of accessing seamlessly
multiple sources are a good match for DoD: we do not need to modify any 
system and can control with fine-granularity how processing is done. We describe
next a few technical details of the ad-hoc engine.

\subsubsection{Materializing a Join Graph}

A join graph contains the information necessary to join a set of N tables. The
join graph nodes represent the tables, and an edge between two nodes indicates that
the two tables can be joined together. Each node contains the attribute on which
the table should be joined. 

To materialize a join graph, joins are performed from the \emph{leaf nodes to inner
ones}. A leaf node is one that is connected to only another one. Every time
a join is performed, its node and edge are removed. The procedure repeats N-1
times, until the resulting joined table is obtained. 

\mypar{On Join Ordering} Choosing the right join ordering is crucial for good
performance; this is one of the classic problems query optimizers aim to
solve. Query optimizers rely on data statistics to estimate what join
orderings are cheaper to compute. Unfortunately, in the scenario we consider, we
cannot assume that the underlying tables will contain statistics, because they
may proceed from different databases, and even file systems. Instead, DoD
records the actual output cardinality as it executes joins, and relies on this
information for future joins, as opposed to use the default \emph{leaf to inner
nodes} join strategy. Since each query view may trigger multiple similar joins,
this strategy is beneficial.

\subsubsection{Expensive-To-Compute Joins}

When materializing join graphs, DoD needs to deal with two practical problems:
expensive joins and joins that have an output larger-than-memory. We explain how
we deal with both problems next:

\mypar{Sampling Joins} Certain joins are so expensive that they become the main
computational bottleneck of the end-to-end view search. Instead of materializing
the entire join, it is possible to join only a \emph{sample} and then run the 4C
method on the sample join. Because 4C runs on a sample, and not the full
join, the view 4C class may change once the full join is materialized.
Therefore, joining only a sample comes with the tradeoff that users may need to
try a few more views before finding the right one.

The main challenge of joining on a sample is selecting the sample so that the
resulting views can be fed into the 4C method. For example, a naive sampling
strategy that selects uniformly from each join graph does not guarantee that
tuples will overlap in the resulting views, because each sampling process is
independent from the others. However, we need samples to overlap, or otherwise,
the 4C method won't find a good classification of views.

To address this problem, DoD employs a strategy that selects samples in a
consistently random way. For every join, it uses a hash function on the join
attribute of the larger table to map the values from all relations to a common
hash space. The sample corresponds to the values with the top-K minimum (or
maximum) hash values, with K indicating the sample size. Because this strategy
is applied consistently across all join graphs, DoD makes sure that the
resulting views can be compared with each other, unlike with independent uniform
sampling. This strategy is reminiscent of conditional random
sampling~\cite{crs}, which has been used before to summarize and compare sets of
elements. As a final note, the sampled views are not guaranteed to contain the
value constraints defined in the input query view. If users want to see the
value constraints in the candidate view, then it's possible to specifically
search for their key in the table and include those as part of the sample,
although it's not necessary for 4C to work. 

\mypar{Larger-than-Memory Joins} Sometimes it is necessary to materialize views
that do not fit in memory, so users can analyze them offline. Because Pandas
assumes that joins can be performed in memory, we implement a spill-to-disk
strategy to overcome this limitation.  However, because the spill-Join is
significantly less efficient than the default in-memory join, we want to use it
only when the join won't fit in memory. The challenge is that without statistics
it is hard to estimate when this is the case.

Our solution is inspired by dynamic query reoptimization~\cite{reoptimization}.
When joining two tables, DoD selects a sample from one table and joins it with
the other table. It measures the output cardinality of the partial join, and
uses it to estimate the output cardinality of the whole table. With a measure of
the memory required by each row, we can then make an informed decision on what
join strategy to use: whether to use spill-Join, our external algorithm
implementation (when it does not fit in memory) or directly in memory (when it
does).

\subsubsection{Caching Optimizations}

There are 3 operations that are the main computational bottlenecks during the
view search performed by DoD: determining whether a candidate group is
joinable, determining whether a joinable group is materializable, and
materializing the group.

We make extensive use of caches to amortize certain operations that are executed 
often. To determine whether a pair of tables can be joined with each other, we
make a call to Aurum. Although this operation is relatively well optimized by
the Aurum engine, it must be performed multiple times, so caching the responses
helps with skipping computation. To determine whether a join graph is
materializable, we need to execute a join query with predicates given by the
cosnstraints in the query view. For a given pair of table-attributes, we can
remember if they do not materialize. Then, before trying to materialize a new
join graph, we first check using this cache that the join graph does not contain
a non-materializable path (this strategy is equivalent to the optimizations
performed by S4~\cite{sfour}). Last, DoD reads tables from data sources in
order to check whether they are materializable and, if so, to materialize them.
Keeping an LRU cache with the read tables helps to avoid expensive IO.

\section{Evaluation}

In this section we evaluate DoD with respect to the two critical metrics of
interest: capacity to reduce the size of the view choice space, and end-to-end
performance. We organize the section around 3 key questions:

\begin{myitemize}
\item Does DoD-4C reduce the size of the view choice space? The default view
choice space size is the number of candidate views generated by a query view. We
want to understand the reduction achieved by 4C.
\item Is the 4C-chasing algorithm fast enough to be practical? Classifying the
candidate views into the 4 categories is an expensive process. The longer time
it takes, the lower the benefits it yields. We compare with a baseline to
demonstrate its efficiency and practicality.
\item Performance of DoD. In this section we conduct experiments to
understand DoD's end-to-end performance, the factors
that contribute to its performance, studying its scalability, as well as the
impact of each of the ad-hoc processing optimizations.
\end{myitemize}

We start the section by presenting the setup and datasets used and then explore
each of the questions above. We conclude the section with a summary of the
results obtained.

\subsection{Setup and Datasets}

We conduct all experiments in a MacBook Pro with 8GB memory and a core i7 with 4
cores and 3GHz speed each. We use two real world datasets in our evaluation:

\textsf{DWH: }The MIT datawarehouse dataset consists of 160 tables that have
been integrated from 116 different databases. The dataset is heterogeneous, with
information concerning many different aspects of the institute, faculty,
facilities, students, subjects, footage, etc. Every table's size in this dataset
is within 100MB.

\textsf{CHE: }This dataset consists of 113 tables from two popular public
chemical databases, ChEMBL~\cite{chembl}, and DrugCentral~\cite{drugcentral}.
The databases contain some overlapping information, but their emphasis is
different, so it is common to conduct integration tasks between them. Tables of
this dataset are of varying sizes, with a few multi-gigabyte tables, others
below 1GB, and others within 100MB.

\smallskip

We use Aurum to build a discovery index for each dataset. We only provide Aurum
with a connector to the databases, and we do not give any information about
PKFKs or similar: all information DoD uses is inferred by Aurum. We use a
collection of 10 query views from the 2 datasets above, 5 queries from each
dataset. Throughout the section, we check that at least one of the candidate
views is a correct view of the DoD output; we have correct views for each of the
10 query views we use.

\subsection{Does DoD-4C reduce human intervention?}

The goal of 4C is to reduce the number of candidate views produced by the search
process. In this experiment we first submit query views to DoD and obtain
the number of candidate views generated. This number is shown in the `Original
Views' column of table~\ref{table:4csummary} (for \textsf{DWH}, we only show the
3 query views that produced a large number of views). This is the size of the
original view choice space: the number of views that users would need to
consider without the 4C method. Using the original candidate views as input,
we use the 4C method and measure the reduction of the view choice space. 
We have implemented two presentation strategies. With \textsf{multi-row}, we
will always obtain 1 multi-row view per schema type, so the more interesting
case, and the one for which we show results, is that of \textsf{4c-summary}. 

The results for \textsf{4c-summary} are shown in the `4c-summary' column of
table~\ref{table:4csummary}. The `x(y)' notation indicates the total number of
views (x), and the number of interactions users must make (y), \ie when choosing
among contradictory views. \textsf{4c-summary} reduces by several factors the
number of views for both datasets. 

Note that users would likely not inspect each of the original candidate views.
Instead, they may select the first candidate view that seems to address their
needs. This, however, may leave a better view uncovered. In contrast, 4C does
not skip any candidate view, leading to a more exhaustive exploration.

\begin{table}[]
\centering
\begin{tabular}{|c|c|c|}
              & \textbf{Original views} & \textbf{4c-summary} \\ \hline
\textbf{DWH2} & 9                       & 4                   \\
\textbf{DWH4} & 99                      & 3(4)                 \\
\textbf{DWH5} & 102                     & 11(11)               \\
\textbf{CHE1} & 12                      & 2(2)                    \\
\textbf{CHE2} & 27                      & 3(4)                 \\
\textbf{CHE3} & 50                      & 14(5)                \\
\textbf{CHE4} & 54                      & 2                   \\
\textbf{CHE5} & 127                     & 23(23)              
\end{tabular}
\caption{Original and reduced views}
\label{table:4csummary}
\end{table}

\smallskip

\noindent\textbf{When does 4C work?} 4C's ability to reduce the size of the choice space
does not depend on the number of original candidate views, but on the actual
content of the views. Consider the column for \textsf{DWH4} and \textsf{DWH5},
as well as \textsf{CHE3} and \textsf{CHE4}. Although in both cases the original
number of views is similar, the reduction is quite different. 

To understand why this is the case, consider a pathological case in which each
view is contradictory with respect to every other view, so every candidate view
must be considered at the output. Or, consider the opposite case where all views
are compatible with each other, and hence summarized into only 1, in which no
candidate view needs to be considered.  Most query views we have used, including
the ones we use in this evaluation, contain a mix of the 4 groups. This is due
to semantic differences and wrong inclusion dependencies. 

\mypar{Note on PKFK} Many candidate views are spurious because they were
obtained using a wrong join path. If the real PKFK-graph was available, the
number of the candidate views would be lower, but the PKFK-graph is rarely
available. An important contribution of DoD is precisely not assuming its
existence and still deal with those spurious views post-hoc.


\subsection{Is 4C-Chasing fast enough?}

4C's ability to reduce the size of the view choice space is only beneficial as
long as the reduction is achieved fast. If users had to wait for hours to
retrieve the results, the benefits would be unclear.

To understand \textsf{4C-Chasing} algorithm's efficiency, we measure its runtime
over the set of query views that yielded a high number of candidate views (we
use those with low candidate views to understand its overhead later). Further,
to justify the design of \textsf{4C-Chasing}, we compare its runtime with a
baseline implementation of 4C, called \textsf{No-Chasing}. \textsf{No-Chasing}
hashes rows so it can classify compatible and contained views quickly, but it
must perform the cell by cell comparison across views to identify complementary
and contradictory ones, as explained in Section~\ref{sec:4c}. A naiver
strategy that inspects each row individually takes too long even when the number
of candidate views is small.

\begin{figure}[h]
  \centering
  \includegraphics[width=0.9\columnwidth]{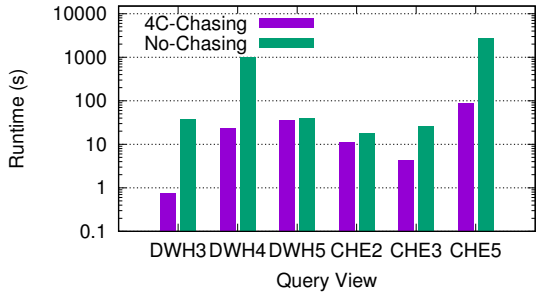}
\caption{Runtime comparison of 4C-Chasing vs No-Chasing for different number of
views}
\label{fig:chasing-benefit}
\end{figure}

The results of \F\ref{fig:chasing-benefit} show that with \textsf{4C-Chasing}
algorithm, we can summarize the views about 2 orders of magnitude faster than
\textsf{No-Chasing} for most queries. Even the more modest improvements of
\textsf{CHE2} and \textsf{CHE3} are of 38\% and 84\% respectively. 

In summary, with \textsf{4C-Chasing} it is possible to vastly reduce the
view choice space in under 2 minutes, making it practical.

\noindent\textbf{What about 4C's overhead?} 
4C is not necessary when the number of candidate views is low, or when
classifying them is very fast, \eg most views are compatible.  In both of these
cases, we want to understand what's the overhead of executing 4C and its impact
on the end-to-end execution. To measure the overhead, we executed
\textsf{4C-Chasing} on query views that either produced few candidate views
(\textsf{DWH1}, \textsf{DWH2}, \textsf{CHE1}) or that produced mostly compatible
views, such as \textsf{CHE4}.

\begin{figure}[h]
  \centering
  \includegraphics[width=0.9\columnwidth]{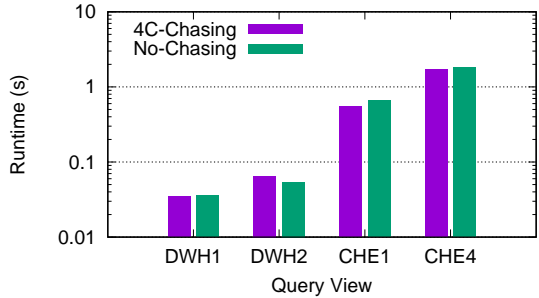}
\caption{Both 4C-Chasing and No-Chasing overhead is negligible when the number
of views is low}
\label{fig:chasing-overhead}
\end{figure}

The results of \F\ref{fig:chasing-overhead} show that both \textsf{4C-Chasing}'s
and \textsf{No-Chasing}'s overhead is low: under 2 seconds in both cases. 
This means that it is possible to always execute 4C because its overhead on the
end-to-end runtime is negligible.

\subsection{Performance of DoD}

The end-to-end performance of DoD does not depend on the 4C method only, but
also on the view search process, which we evaluate here. For that, we measure
the time it takes for DoD to generate all the candidate views from the input
queries of both datasets. In the case of \textsf{DWH}, we fully materialize the
views, while in the case of \textsf{CHE}, we use the consistent sampling
strategy (see Section~\ref{sec:engine}), because the underlying tables of the
dataset are much larger and many joins would run out of memory otherwise.

\begin{table}[]
\centering
\begin{tabular}{|c|c|c|c|c|c|c|}
              & \textbf{\begin{tabular}[c]{@{}c@{}}Runtime\\ (s)\end{tabular}} &
\textbf{\begin{tabular}[c]{@{}c@{}}CG\end{tabular}} &
\textbf{\begin{tabular}[c]{@{}c@{}}P\end{tabular}} &
\textbf{\begin{tabular}[c]{@{}c@{}}JG\end{tabular}} &
\textbf{\begin{tabular}[c]{@{}c@{}}Join\\ graphs\end{tabular}} &
\textbf{\begin{tabular}[c]{@{}c@{}}MG\end{tabular}} \\ \hline
\textbf{DWH1} & 0.27                                                           &
3 & 1                                                                   & 0
& 0                                                              & 2
\\
\textbf{DWH2} & 46.8                                                           &
22 & 48                                                                 & 9
& 2140                                                           & 2
\\
\textbf{DWH3} & 4.6                                                            &
15  & 9                                                                & 3
& 9                                                              & 9
\\
\textbf{DWH4} & 139.1                                                          &
15 & 12                                                                & 12
& 572                                                            & 99
\\
\textbf{DWH5} & 683                                                            &
24 & 47                                                                 & 12
& 2148                                                           & 102
\\
\textbf{CHE1} & 856                                                            &
4  & 2                                                                 & 2
& 69                                                             & 12
\\
\textbf{CHE2} & 15.9                                                           &
9  & 14                                                                 & 2
& 38                                                             & 27
\\
\textbf{CHE3} & 16.1                                                           &
11  & 6                                                                & 5
& 61                                                             & 50
\\
\textbf{CHE4} & 249.1                                                          &
5  & 2                                                                 & 2
& 79                                                             & 54
\\
\textbf{CHE5} & 37.5                                                           &
11 & 11                                                               & 6
& 127                                                            & 122                                                                     
\end{tabular}
\caption{Statistics about e2e view search process for both datasets. Candidate
groups (CG), Pairs of tables (P), Joinable groups (JG), Materializable groups (MG)}
\label{table:e2edod}
\end{table}

\begin{figure}[h]
  \centering
  \includegraphics[width=0.9\columnwidth]{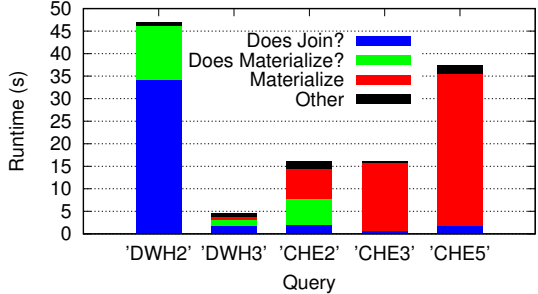}
\caption{Total runtime split by component and for different queries (1/2)}
\label{fig:e2edod}
\end{figure}

\begin{figure}[h]
  \centering
  \includegraphics[width=0.9\columnwidth]{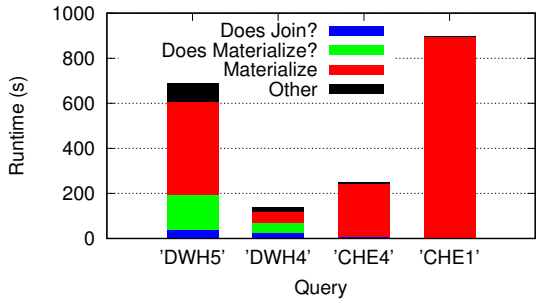}
\caption{Total runtime split by component and for different queries of (2/2)}
\label{fig:e2edod_chem}
\end{figure}

Table~\ref{table:e2edod} shows the end to end runtime (first column) of DoD for
different queries (shown in the rows) as well as statistics about each
execution. The total runtime depends on several aspects of the search process:
the effort to understand if tables join (\textsf{Does Join?}), the effort to
check whether the joinable tables are materializable (\textsf{Does
Materialize?}), and the time to actually materialize views
(\textsf{Materialize}). Runtime numbers, per operation are shown in
\F\ref{fig:e2edod} and \F\ref{fig:e2edod_chem}. The figures also include a
category \textsf{Other}, to represent time spent doing work that does not fit in
the categories above. We analyze each cost next:

\smallskip



\noindent\textbf{Does Join?} The \textsf{P} column in the table indicates 
the pairs of tables for which the system must find join paths (see
Section~\ref{subsubsec:joinable}). This number depends on the number of
candidate groups, as well as the number of tables per group. The higher the
number of pairs of tables, the costlier the \textsf{Does Join?} operation is, as
confirmed by the results in the table and figures. This effect is more clearly
seen in the case of the \textsf{DWH} dataset, because there are more pairs of
tables for which to find join paths. 

\smallskip

\noindent\textbf{Does Materialize?} The \textsf{Join Graphs} column of
Table~\ref{table:e2edod} indicates how many potential ways we can join the
pairs of tables identified before.  This number depends to some extent on the
number of joinable groups found (column \textsf{JG}), but mostly on how
connected the specific tables are within Aurum's discovery index. For each
joinable graph, DoD must check if the graph is materializable (see
Section~\ref{subsubsec:ismaterializable}), so naturally, the higher the number
of join graphs, the higher the cost of \textsf{Does Materialize?}. 

\smallskip

\mypar{Materialize} Finally, the \textsf{MG} column (Table~\ref{table:e2edod}),
materializable groups, indicates how many of the materializable join graphs, if
materialized, would lead to a view that satisfies the input query view. This
variable helps with understanding how much of the runtime is dedicated to
actually materialize the joins to produce the output candidate views. In turn,
the cost of materialization depends on how many views to materialize, and how
expensive each join is. For example, in the case of the \textsf{DWH} dataset,
query \textsf{DWH5} is dominated by the materialization time. Each join is quite
cheap, but there are 102 joins to perform, which explains the relatively high
runtime. In the case of the \textsf{CHE} dataset, runtime is dominated by the
materialization step. This is using using the sampling strategy of
Section~\ref{sec:engine}, which is necessary for this dataset because the
underlying tables are large. Without it, many of the joins run out of memory,
and those that don't, take (each) about 6-8x longer to complete. The reason for
the bottleneck is that in order to obtain the sample, it is necessary to read
the entire relation into memory first---Pandas does not allow to efficiently
read a relation selectively.

\subsubsection{Scalability}

DoD is built on top of Aurum, and benefits from its scalability~\cite{aurum}.
However, DoD uses other operations, such as those performed by the ad-hoc query
engine to understand if a join graph is materializable, and to materialize it
when it is. In this experiment, we want to understand whether DoD's view search
scales as the amount of work to perform grows. Ideally, the runtime depends
linearly on the amount of work that is necessary to perform. 

\begin{table}[h]
\centering
\begin{tabular}{|l|c|c|c|}
\multicolumn{1}{|c|}{}            & \textbf{\begin{tabular}[c]{@{}c@{}}X1\\
(9)\end{tabular}} & \textbf{\begin{tabular}[c]{@{}c@{}}X2\\ (36)\end{tabular}}
& \textbf{\begin{tabular}[c]{@{}c@{}}X4\\ (192)\end{tabular}} \\ 
\hline
\textbf{\# Pairs tables}       
& 5 & 38 & 172 \\ 
\textbf{\# Join graphs}
& 9 & 72  & 576 \\ 
\textbf{\# Materializable groups} 
& 3 & 24 & 192 \\ 
\end{tabular}
\caption{Statistics for the scalability experiment}
\label{table:scalability}
\end{table}

In order to conduct this experiment, we use the \textsf{DWH2} query view as
input, however, we use 3 instances of the problem with different scale factors.
The first one uses the original \textsf{DWH} dataset. The second one duplicates
the original (X2), and the third one quadruplicates it (X4). Note that we are
increasing the complexity exponentially, not linearly, as we are multiplying the
number of tables to check for joinability, as well as materialization, and the
total number of candidate views to materialize. The exact numbers that indicate
the amount of work are shown in table~\ref{table:scalability}. 

We measure the time the system takes to perform each of these operations, and
compute the `Normalized Time' as the time per operation, \ie instead of the
aggregated one. A scalable system would show that the time per operation remains
constant regardless the scale factor.

\begin{figure}[h]
  \centering
  \includegraphics[width=0.95\columnwidth]{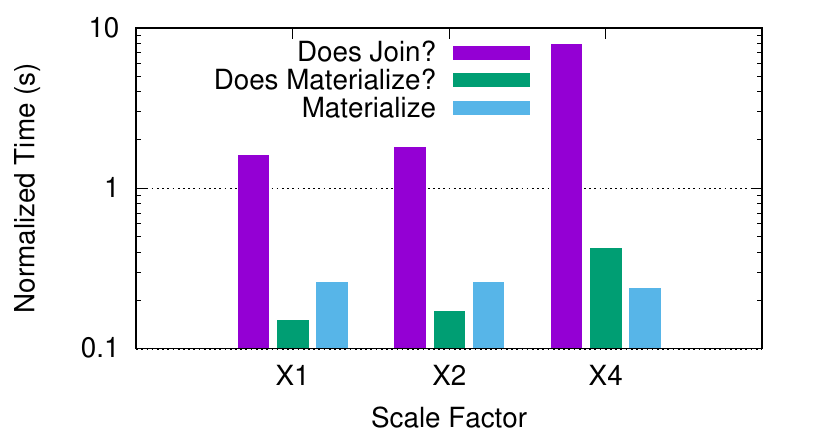}
\caption{Scalability of main time-consuming procedures of DoD as the scale
factor increases: two times (X2), and four times (X4).}
\label{fig:scalability}
\end{figure}

\F\ref{fig:scalability} shows the results of the experiment. The \textsf{X} axis
shows the scale factor and the \textsf{Y} shows the normalized time (time per
operation). The figure shows that the time for \textsf{Does Materialize?} and
\textsf{Materialize} remains constant despite the work growth due to the
increase of the scale factor. The \textsf{Does Join?} operation cost increases a
bit for the 4X case, but it remains within the same scale factor.

\subsubsection{Ablation test}

In this last section, we conduct experiments to understand the benefits of the
optimizations of the ad-hoc query engine. For that, we select two
query views, \textsf{DWH4} and \textsf{DWH2}. We select these two because they
perform different amount of work and produce different number
of views. We execute the query views with different optimizations enabled and
measure their runtime.

\begin{figure}[h]
  \centering
  \includegraphics[width=0.9\columnwidth]{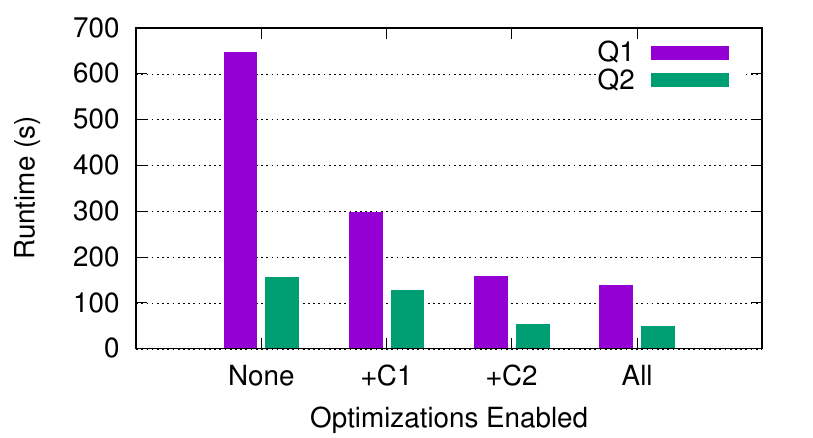}
\caption{Performance improves from left to right as more optimizations are
enabled (2 different queries)}
\label{fig:ablation}
\end{figure}

\F\ref{fig:ablation} shows the results of executing the queries without any
optimization enabled, \textsf{None} in the \textsf{X} axis. We then activate
optimizations incrementally. First, we activate the caching mechanism for the
\textsf{Does Join?} operation (\textsf{+C1} in the Figure), then the \textsf{Does
Materialize?} optimization, \textsf{+C2}, and last we execute the full
DoD, labeled \textsf{All}, which includes the relation cache to avoid reads from
disk.

The figure clearly shows the performance benefits of \textsf{+C1} and
\textsf{+C2}. The advantages of the optimization for the \textsf{Materialize}
operation, which are the difference between the right-most point in the figure
and the previous one, are less prominent. The relative value of each of these
optimizations depends on the amount of work assigned to each of these
operations during the view search. However, in aggregate, the optimizations
implemented reduce the time with respect to a baseline implementation a factor
of 4-6X.

\mypar{Note on sampling join} Without sampling join, many of the joins on large
tables run out of memory. Those that do not run out of memory take 6-8X more
time to complete.

\subsection{Summary of Results}

The evaluation focused on two key metrics of interest. First, we showed how 4C
reduces the size of the view choice space by an order of magnitude. It achieves
this reduction while executing within a few minutes at most.  End-to-end, DoD
finds candidate views within minutes, and, in some cases within a few seconds.
In essence, DoD trades expensive human time for cheaper compute cycles to
accelerate view search and presentation.

\section{Related Work}

In this section, we explain how DoD fits into the extensive literature on
data integration.

\mypar{View-by-Example} The initial ideas of this approach were presented in
\cite{schemamappingasquerydiscovery}, in a theoretical way, and later
implemented in Clio~\cite{clio}. More recently, on the theoretical side, new
contributions have helped understand the problem in more
depth~\cite{approxmappings}.  The current class of view-by-example systems
\cite{samples4, sfour, mweaver, exemplarqueries} aim to find views for users using
a view definition, like DoD. Unlike DoD, these systems make two key assumptions
that sets them apart: i) assume knowledge about how to join any pair of tables;
ii) assume the total views that satisfy the input definition is small. Although
the more advanced systems have a way of ranking the resulting views, such as in
S4~\cite{sfour}, this does not help disambiguate semantically different views. In
contrast, DoD does not make these assumptions and deals with these
problems directly.

\mypar{Data Integration} In the early years, data integration required humans
to provide mappings (more in general, a mediated schema) between heterogeneous
sources in order to understand how to combine them \cite{clio, teenageyears,
manifold}. As the number of sources has grown and become more heterogeneous,
these intensive human-driven approaches have led to more automatic
alternatives. DoD does not require users to provide any mediated schema or
mapping between sources. Instead, users only participate at the end to select a
view among a reduced choice space. Furthermore, with the \textsf{multi-row}
presentation strategy, human involvement is further reduced.

\mypar{Aiding Mapping Selection} Because creating mappings for integration is
hard, many approaches appeared to assist users with the selection of the best
mapping. In \cite{datadrivenunderstanding}, this problem receives a theoretical
treatment. Later, practical approaches to drive the mapping selection have
appeared \cite{bonifati}. In these approaches, the mapping selection is driven
by the user, who is still in charge of determining what mappings are correct.
These approaches are appropriate when users are assumed to be familiar with the
schema. DoD does not assume knowledge of the schema from users, and does not ask
them to reason about mappings. With DoD, users can still inspect the mappings
and make decisions at that level, but they also have the actual view created,
which they can use to make the final decision.

\mypar{Data Discovery} Data discovery solutions such as Goods \cite{goods} and
Infogather~\cite{infogather} assist with identifying relations of interest, but
they do not solve the end to end problem of view creation. Data catalog
solutions~\cite{catalog1, catalog2, catalog3} help organizing, annotating, and
sharing data sources but do not directly help with finding or joining the
datasets. Aurum~\cite{aurum} exposes a data discovery API that permits answering
many different queries, as long as analysts know how to write code, which is
often not the case. 

\mypar{Keyword search in databases} Keyword search systems \cite{discover,banks,
keyword1} are precursors of many discovery and view-by-example systems today.
These systems aim to identify tuples in databases that contain the input
keywords. To find such tuples, the systems identify ways to join the involved
tables together. However, these systems do not take attribute names as input,
they assume the existence of a PKFK-graph, and they do not help with assembling
the full view, nor presenting the many views in a concise way.


\section{Conclusion}

DoD is a system that identifies views as combinations of multiple tables that
may span multiple heterogeneous data sources. DoD only requires a query view as
input, and does not assume the existence of a PKFK-graph. With the query view,
it finds all views that fully or partially satisfy it, and then reduces the size
of the choice space using the \textsf{4c-chasing} algorithm and a presentation
strategy. The 4C method is an effective way of dealing with views that are
semantically different, as well as with data quality problems.


\bibliographystyle{abbrv}
\bibliography{main}
\balance

\end{document}